\documentclass[amssymb,amsmath,aps,showpacs,nofootinbib]{revtex4}
\usepackage[]{graphicx}
\usepackage[]{amsmath}

\def\ket#1{| #1 \rangle}
\def\bra#1{\langle #1 |}

\def\kb#1#2{| #1 \rangle\!\langle #2 |}
\def\II{1\!\mathrm{l}}
\def\cA{\mathcal{A}}
\def\cB{\mathcal{B}}
\def\cC{\mathcal{C}}
\def\cE{\mathcal{E}}
\def\cG{\mathcal{G}}
\def\cH{\mathcal{H}}
\def\cK{\mathcal{K}}
\def\cM{\mathcal{M}}
\def\cN{\mathcal{N}}
\def\cS{\mathcal{S}}
\def\cT{\mathcal{T}}

\def\bC{\mathbb{C}}

\def\up{\!\!\uparrow}
\def\dn{\!\!\downarrow}
\def\upp{\uparrow\!\!}
\def\dnn{\downarrow\!\!}

\begin{document}

\title{Toy Model for a Relational Formulation of Quantum Theory}
\author{David Poulin}
\email{dpoulin@iqc.ca}
\affiliation{School of Physical Sciences, The University of Queensland, QLD 4072, Australia}

\date{\today}

\begin{abstract}
In the absence of an external frame of reference --- i.e. in background independent theories such as general relativity --- physical degrees of freedom must describe {\em relations} between systems. Using a simple model, we investigate how such a relational quantum theory naturally arises by promoting reference systems to the status of dynamical entities. Our goal is to demonstrate using elementary quantum theory how any quantum mechanical experiment admits a purely {\em relational} description at a fundamental level, from which the original ``non-relational" theory emerges in a semi-classical limit. According to this thesis, the non-relational theory is therefore an approximation of the fundamental relational theory. We propose four simple rules that can be used to translate an ``orthodox" quantum mechanical description into a relational description, independent of an external spacial reference frame or clock. The techniques used to construct these relational theories are motivated by a Bayesian approach to quantum mechanics, and rely on the noiseless subsystem method of quantum information science used to protect quantum states against undesired noise. The relational theory naturally predicts a fundamental decoherence mechanism, so an arrow of time emerges from a time-symmetric theory. Moreover, there is no need for a ``collapse of the wave packet" in our model: the probability interpretation is only applied to diagonal density operators. Finally, the physical states of the relational theory can be described in terms of ``spin networks" introduced by Penrose as a combinatorial description of geometry, and widely studied in the loop formulation of quantum gravity. Thus, our simple bottom-up approach (starting from the semi-classical limit to derive the fully relational quantum theory) may offer interesting insights on the low energy limit of quantum gravity. 
\end{abstract}

\pacs{03.67.-a,04.60.Pp,03.65.Yz}

\maketitle

\section{Introduction}\label{intro} 

To combine the two main physical theories of the twentieth century --- quantum mechanics and general relativity --- it is important to clearly identify the chief insights they offer on the physical world. Quantum mechanics establishes a mathematical apparatus --- Hilbert space, canonical quantization, etc. --- that sets a general framework to describe physical systems. Here, we will assume that this general framework is essentially correct. The main lesson retained from general relativity is that physical theories should {\em not} be formulated in terms of a background reference frame, but rather should be {\em relational}; a point of view emphasized by Rovelli \cite{Rov96a,Rov04a} among other. Starting from plain elementary quantum mechanics, we investigate consequences of background independence.

More precisely, we will argue that an orthodox (non-relational) physical description can be made purely relational by applying the following four simple rules. 
\begin{enumerate}
\item Treat everything quantum mechanically.
\item Use Hamiltonians with appropriate symmetries.
\item Introduce equivalence classes between quantum states related by an element of the symmetry group.
\item Interpret diagonal density operators as probability distributions.
\end{enumerate}
In the appropriate semi-classical limits, the relational theory will be equivalent to the orthodox quantum description, in the sense that it leads to the same physical predictions. However, away from these limiting regimes, the relational theory also predicts new phenomenon, such as a fundamental decoherence process~\cite{GPP04a}. 

Although these rules seem reasonable to us, there is no way in which they are fundamentally ``right". We will nevertheless take them for granted and see where they take us. One is then free to like or dislike our conclusions, until the new predictions can be tested experimentally. Rule 1 states that we should not resort to semi-classical approximations in the description of a physical system. If necessary, these approximation can be used to perform computations at a later stage, but should not appear in the fundamental formulation. These approximations are often responsible for breaking the symmetries of the system, and in effect, introduce a background reference frame. For example, treating an external magnetic field as classical provides a natural axis to quantize angular momentum. Note that quantum theory is not required to arrive at a relational description: there are perfectly valid relational classical theories. In a classical relational theory, it is often possible to  choose an arbitrarily system as a reference frame and recover a non-relational theory, e.g. by working in the rest frame of a specific particle. In a quantum settings however, switching from the relational to the non-relational description will always require some sort of {\em approximation} since reference frames defined with respect to quantum systems are subject to quantum fluctuations \cite{AK84a,Rov91b,Tol97a,Maz00a}. These quantum effects can be made arbitrarily small by increasing the mass of the reference system. However, in practical situations reference systems have finite masses. Moreover, increasing their masses will induce gravitational distortions to the measured quantities \cite{Wig57b,GLM04a}, so it is essential to keep the masses finite. Here, we will neglect gravitational corrections and focus on quantum mechanical effects.

Rule 1 thus implies that fundamental descriptions must be relational. Hence, this rule is intimately related to Rule 2. Once every physical system --- including reference systems --- is treated quantum mechanically, there should be nothing left to break the fundamental symmetries, and accordingly, the Hamiltonians should also have the appropriate symmetries. 

Rule 3 is used to get rid of unphysical information in the description of the system. States that are related by a transformation that belongs to the symmetry group of the system should be regarded as physically equivalent. This Rule is quite natural if one adopts a Bayesian interpretation of quantum states~\cite{CFS02a,Fuc02b}. Following a Bayesian prescription, the lack of an external reference frame leads to group averaging of the quantum states. The effect of this group average will be to randomize the unphysical degrees of freedoms --- those defined with respect to an external reference frame --- while leaving the relational (hence physical) ones unchanged. At this stage, the unphysical degrees of freedom can be removed from the description as they carry no information. This procedure is inspired by the noiseless subsystems method~\cite{KLV00a,Zan01b,KBLW01a} used to protect the state of ``virtual" quantum systems in quantum information science. In this language, the physical degrees of freedom form noiseless subsystems of a noise algebra, where the noise operators are elements of the symmetry group of the system. At a formal level, Rule 3 says that we must quotient the state space of physical systems by their symmetry group. We note that the Bayesian prescription differs from the ``coherent" group average commonly encountered in quantum gravity (see \cite{Mar00a} and references therein), leading to distinct physical descriptions. 

Finally, Rule 4 gives the standard interpretation to quantum states, but circumvents the problematic collapse of the wave packet. All four Rules will get clarified as we apply them to a simple example in the next Section.

The outline of the paper is as follows. In the next Section, we will present the simplest textbook quantum mechanical system: a spin-$\frac 12$ particle immersed in a uniform constant magnetic field. Despite its simplicity, the orthodox description of this system violates all four Rules, so it provides a good starting point to illustrate our procedure. In the following Sections (\ref{sec:measure}, \ref{sec:dynamics}, and \ref{sec:time}), we gradually apply our four rules to this system and arrive at a relational theory that is equivalent to the original theory in the appropriate ``macroscopic" limit. Away from these limits however the relational theory predicts new phenomenon such as a fundamental decoherence mechanism, that can clearly be seen in our numerical analysis. In Section~\ref{sec:nns}, we describe the general picture that derives from our four Rules. The main ingredients of the construction are the noiseless subsystems of quantum information science that will be briefly described. Section~\ref{sec:decoherence} discusses the fundamental decoherence mechanism that arises from the relational theory. In short, decoherence will occur whenever clocks are of finite size and therefore subject to quantum fluctuations~\cite{GPP04a}, leading to an arrow of time in a time-independent theory. Section~\ref{sec:SN} establishes a connexion between the basis states of the relational theory and ``spin networks"  introduced by Penrose~\cite{Pen71a} as a combinatorial description of geometry and widely studied in the loop formulation of quantum gravity~\cite{RS95a, Baez95a} (see also \cite{Majo00a}). It is our hope that the slight distinction between how spin networks arise in our ``semi-classically inspired" model and how they are used in loop quantum gravity will yield some new insights on the low energy regime of quantum gravity.
In Section~\ref{sec:other_symmetries}, we speculate about possibles extensions of the program. Finally, we conclude with a summary in Section~\ref{sec:conslusion}. 

\section{An example}
\label{sec:example}

We begin by illustrating our program with the simplest quantum mechanical system: the system ($\cS$), a spin-$\frac 12$ particle, is interacting with a uniform magnetic field $\vec{B}$. In this toy Universe, the spacetime manifold has the topology $S_2 \times S_1$; there are only orientations in 3-space (hence the 2-sphere) and time (which we assume takes a finite range with periodic boundary conditions, and hence has the topology of a circle). Accordingly, the fundamental symmetries are $SO(3)$ and $U(1)$. The orthodox description of this system goes as follows. Without loss of generality, we assume that $\vec{B}$ is along the $x$ axis, so the system's Hamiltonian is $H^\cS = -B\sigma_x^\cS$ where all physical constants are absorbed in $B$. The system's initial conditions are specified by the state $\ket{\psi(0)}^\cS = \alpha\ket{\up}^\cS + \beta\ket{\dn}^\cS$, where the quantization axis is arbitrarily chosen to be along the $z$ direction (this will be the case throughout this manuscript, unless specified otherwise). At time $t$, the state of the system is 
\begin{subequations}
\begin{eqnarray}
\ket{\psi(t)}^\cS &=& \alpha(t)\ket{\up}^\cS + \beta(t)\ket{\dn}^\cS,\\
\alpha(t)&=& \alpha\cos(Bt/2) +i \beta\sin(Bt/2) \\
\beta(t)&=& i\alpha\sin(Bt/2) + \beta\cos(Bt/2) .
\end{eqnarray}
\label{eq:time_coeff}
\end{subequations}

If we wish to measure the value of the spin of the system at time $\tau$, say along the $z$ axis, we must introduce a measurement apparatus $\cA$ that couples to $\cS$. The time-dependent interaction Hamiltonian 
\begin{equation}
H^{\cS\cA}(t) = -g\delta(t-\tau)\sigma^\cS_z\otimes\sigma^\cA_y
\label{eq:coupling}
\end{equation}
is a good choice of ``measurement" coupling. The coupling constant is set to $g=2\pi$. Given the initial state of the apparatus $(\ket\up^\cA + \ket\dn^\cA)/\sqrt 2$, the joint state of $\cS-\cA$ at time $\tau_+$ immediately after the interaction 
is 
\begin{equation*}
\ket{\Psi(\tau_+)}^{\cS\cA} = \alpha(\tau)\ket\up^\cS\otimes\ket\up^\cA + \beta(\tau)\ket\dn^\cS\otimes\ket\dn^\cA.
\end{equation*}
This pre-measurement phase establishes correlations between $\cS$ and $\cA$. The next step of the measurement process is the collapse of the wave function, which asserts that the measurement apparatus, being a ``classical" object, cannot be in a quantum superposition, so it rapidly collapses into either the up or down state, each with probabilities given by amplitude squared. In certain circumstances, this step can be given an operational justification~\cite{Zur03a,OPZ04a,Pou04a}. After this stage, the pair $\cS-\cA$ is described by the mixed state 
\begin{equation*}
\rho^{\cS\cA} = |\alpha(\tau)|^2\kb\up\upp^\cS \otimes\kb\up\upp^\cA
 + |\beta(\tau)|^2\kb\dn\dnn^\cS \otimes\kb\dn\dnn^\cA.
\end{equation*}
The interpretation of this state is that both $\cS$ and $\cA$ are either in the up state with probability $|\alpha(\tau)|^2$, or both in a down state with probability $|\beta(\tau)|^2$, given by Eq.~(\ref{eq:time_coeff}), and this completes the measurement process.

This description is obviously not background independent as it makes explicit use of an external coordinate system. In the case of the external field, this dependence is explicit: $\vec B \propto \hat x$, where $\hat x$ makes reference to a coordinate system. The dependence of the measurement apparatus on an external reference frame is twofold. First, the coupling Hamiltonian Eq.~(\ref{eq:coupling}) used to establish correlation between system and apparatus is neither rotationally or time-translational invariant, in violation of Rule 2. Moreover, the collapse phase requires the specification of a preferred observable: classical objects cannot be in superposition involving different values of this preferred observable. We say the preferred observable is {\em superselected}. In the above example, the preferred observable was the angular momentum along the $z$ axis, once again making reference to an external coordinate system. 

We will refer to the example presented in this section as the ``toy model". In the next subsections, we will apply our four Rules to the toy model and eliminate the need for an external reference frame,  demonstrating how one naturally arrives at a purely relational theory. Before doing so, we must pause to establish some notation. In what follows, we use several particle with different angular momentum (or spin) to describe the toy model. To avoid confusion, we adopt the following notation. Each particle is given a name that is represented by a calligraphic capital letter, e.g. $\cA$. Operators, states, and Hilbert space referring to this particle will have the associated letter as a superscript. The quantum number associated to the total angular momentum of the particle is represented by the same capital letter in roman fonts, while the quantum number for the $z$ component will be labeled by the lower case letter. For particle $\cA$, this gives $(J^\cA)^2\ket{A,a}^\cA = A(A+1)\ket{A,a}^\cA$ and $J^\cA_z\ket{A,a}^\cA = a\ket{A,a}^\cA$, with $\ket{A,a}^\cA \in \cH^\cA = \mathbb C^{2A+1}$. Exception will be made for spin-$\frac 12$ particles where the up and down arrows are used. Finally, we denote $\cB(\cH)$ the set of bounded linear operators acting on $\cH$.

\subsection{Measurement}
\label{sec:measure}

Our goal is now to demonstrate how our four Rules naturally lead to a background independent relational theory that is equivalent to the orthodox description {\em in the appropriate limits}. Hence, in this section and the in following, we will often be interested in various limiting regimes of the relational theory. These limits are not constitutive to the theory: their sole purpose is to demonstrate compatibility with known regimes. Of course, the new and interesting physics will arise when the relational theory is analyzed away from these limits. 

Let us first assume that there is no system Hamiltonian, so the system's state is $\alpha\ket\up^\cS + \beta\ket\dn^\cS$ at all times. To perform a spin measurement in the absence of an external reference frame, we need a {\em gyroscope} $\cG$. Following Rule 1, we should treat this gyroscope quantum mechanically. A good choice consists of a spin-$G$ particle\footnote{Note that the gyroscope can be a composite particle. For example, the gyroscope could be a ferromagnet composed of $10^{23}$ spin-$\frac 12$ particles all roughly aligned in the same direction, which for our purposes behaves as a single particle with large spin.} with large value of $G$, prepared in a state of maximal angular momentum along the $z$ direction, i.e. in the quantum state $\ket{G,g=G}^\cG$, which we abbreviate $\ket{G,G}^\cG$. States of maximum angular momentum along a certain axis, also called $SU(2)$ coherent states, are appreciated for their semi-classical properties, and as such, they will be used extensively here to recover the non-relational limit. Note that the description of a gyroscope relies on an external coordinate system, but we will soon get rid of it. 

At any given time, the joint state of the system and  the gyroscope is thus 
\begin{equation}
\ket{\Psi}^{\cS\cG} = (\alpha\ket\up^\cS + \beta\ket\dn^\cS)\otimes \ket{G,G}^\cG.
\label{eq:state_SG}
\end{equation}
As noted above, this state describes unphysical degrees of freedom as it is defined relative to an non-existing coordinate system. To eliminate this pathology, we follow Rule 3 and introduce equivalence classes between states in the composite Hilbert space $\cH^{\cS\cG} = \bC^2\otimes\bC^{2G+1}$. For this, we represent the quantum state of Eq.~(\ref{eq:state_SG}) by the density operator $\rho^{\cS\cG} = \kb\Psi\Psi^{\cS\cG} \in \cB(\cH^{\cS\cG})$. The equivalence classes are obtained by applying the trace preserving completely positive (TPCP) map $\cE^{\cS\cG}:\cB(\cH^\cS\otimes\cH^\cG) \rightarrow \cB(\cH^\cS\otimes \cH^\cG)$ defined by the action
\begin{equation}
\cE^{\cS\cG}(\rho) = \int_{SO(3)} R^{\cS\cG}(\Omega) \rho R^{\cS\cG}(\Omega)^\dagger d\Omega,
\label{eq:CPmap}
\end{equation}
where $R^{\cS\cG} = R^\cS\otimes R^\cG$ is the unitary representation of the rotation group on the pair $\cS-\cG$, and $d\Omega$ is the invariant Haar measure on $SO(3)$. The resulting state $\rho^{\prime\cS\cG} = \cE^{\cS\cG}(\rho^{\cS\cG})$ is thus rotationally invariant. The map $\cE$ generalizes to any number of particles in an obvious manner.

The representation $R^{\cS\cG}$ is generated by the {\em total} angular momentum operator $\vec J^{\cS\cG} = \vec \sigma^\cS + \vec J^\cG$, where $\vec \sigma^\cS = (\sigma_x^\cS, \sigma_y^\cS, \sigma_z^\cS)$ and $\vec J^\cG = (J_x^\cG,J_y^\cG,J_z^\cG)$ are the system's and gyroscope's angular momentum operators respectively. The representation therefore commutes with the operators $(J^{\cS\cG})^2 = \vec J^{\cS\cG}\cdot\vec J^{\cS\cG}$, $(\sigma^\cS)^2$ and $(J^\cG)^2$. Hence, to study the effect of $\cE^{\cS\cG}$, it is useful to express $\ket\Psi^{\cS\cG}$ in terms of the total angular momentum:
\begin{equation*}
\ket\Psi^{\cS\cG} = \alpha\ket{G+\tfrac 12,G+\tfrac 12;\tfrac 12;G} 
+ \frac{\beta}{\sqrt{2G+1}}\ket{G+\tfrac 12 ,G-\tfrac 12;\tfrac 12;G} 
+ \frac{\beta\sqrt{2G}}{\sqrt{2G+1}} \ket{G-\tfrac 12, G-\tfrac 12; \tfrac 12;G}.
\end{equation*}
Above, we use standard angular momentum notation: $\ket{j,m;j_1;j_2}$ is a simultaneous eigenstate of $(J^{\cS\cG})^2$, $J^{\cS\cG}_z$, $(\sigma^\cS)^2$, and $(J^\cG)^2$ with eigenvalues $j(j+1)$, $m$, $j_1(j_1+1)$, and $j_2(j_2+1)$ respectively (see e.g. \cite{Sak94a}). As $J^{\cS\cG}_z$ is the only operator defined with respect to the external reference frame, the effect of $\cE$ on this state can be readily anticipated: it randomizes the value of $m$ while leaving the other quantum numbers unaffected~\cite{BRS0b}. Indeed, the expression we get  is
\begin{equation*}
 \cE^{\cS\cG}(\rho^{\cS\cG}) = \left[|\alpha|^2 + \frac{|\beta|^2}{2G+1}\right] \kb{G+\tfrac 12;\tfrac 12;G}{G+\tfrac 12;\tfrac 12;G} \otimes \frac{\II_{2G+2}}{2G+2}
+ \frac{2G|\beta|^2}{2G+G} \kb{G-\tfrac 12;\tfrac 12;G}{G-\tfrac 12;\tfrac 12;G} \otimes \frac{\II_{2G}}{2G},
\end{equation*}
where the identity operators $\II$ act on the $J^{\cS\cG}_z$ sectors. Thus, the unphysical degree of freedom associated with $J^{tot}_z$ is now in a maximally mixed state and can be removed from the physical description to arrive at
\begin{equation}
\rho^{\cS\cG}_{\rm physical} = \left[|\alpha|^2 + \frac{|\beta|^2}{2G+1}\right] \kb{G+\tfrac 12;\tfrac 12;G}{G+\tfrac 12;\tfrac 12;G} 
+ \frac{2G|\beta|^2}{2G+1} \kb{G-\tfrac 12;\tfrac 12;G}{G-\tfrac 12;\tfrac 12;G}.
\label{eq:rho_phys}
\end{equation}
Rule 4 gives the desired interpretation to this state: a spin-$G$ and a spin-$\frac 12$ particle (here the gyroscope and the system respectively) are either parallel, yielding a total angular momentum $G+\frac 12$, or antiparallel, yielding $G-\frac 12$. The probabilities associated to these two alternatives are $P(\mathrm{parallel}) = |\alpha|^2 + |\beta|^2/(2G+1)$ and $P(\mathrm{antiparallel}) = 2G|\beta|^2/(2G+1)$. When the gyroscope is of a macroscopic size, we recover the familiar probabilities $|\alpha|^2$ and $|\beta|^2$.  Note that these probabilities are the diagonal entries of a diagonal density matrix: no collapse was required to recover the probability rule. 

The general picture illustrated here for a spin-$\frac 12$ particle holds for an arbitrary spin. This follows from the fact that the Clebsch-Gordan coefficients $C(S,s;G,g;J,j) = (\bra{S,s}\otimes\bra{G,g})\ket{J,m;S,G}$ satisfy\footnote{We have verified this limit numerically up to accuracy roughly 1\% for $G$ of a few hundreds, but have not been able to derive it analytically.}
\begin{equation}
\lim_{G\rightarrow\infty} \big|C(S,s;G,G;G+s+\Delta,G+s)\big|^2 = 
\left\{
\begin{array}{l}
1 \ \ \mathrm{ if }\  \Delta = 0 \\
0 \ \ \mathrm{otherwise}.
\end{array}
\right.
\label{eq:limitCG}
\end{equation}
Thus, given a spin-$S$ system in state $\ket{\psi}^\cS = \sum_{s} \alpha_{s}\ket{S,s}$ and a gyroscope in state $\ket{G,G}^\cG$, we can write the combined state in terms of the total angular momentum as
\begin{equation*}
\ket{\Psi}^{\cS\cG} = 
\sum_{s,J}  \alpha_s C(S,s;G,G;J,G+s) \ket{J,G+s;S;G}.
\end{equation*}
When the gyroscope reaches macroscopic scales, this state approaches $\sum_s \alpha_s \ket{G+s,G+s;S;G}$ by virtue of Eq.~(\ref{eq:limitCG}). Rule 3 thus implies
\begin{equation}
\rho^{\cS\cG}_{\mathrm{physical}} \approx \sum_s |\alpha_s|^2 \kb{G+s;S;G}{G+s;S;G}.
\end{equation}
This state has the desired interpretation: the angle $\theta$ between the gyroscope and the system satisfies $\cos(\theta) = s/S$ with probability $|\alpha_s|^2$. 

\subsection{Dynamics}
\label{sec:dynamics}

By promoting the reference frame to the status of a dynamical entity, a gyroscope, we have demonstrated how the quantum probability rule is recovered in a macroscopic limit of a background independent theory (see also e.g. \cite{AK84a,Rov91b,Tol97a,Maz00a} and references therein).  The next step is to introduce non-trivial dynamics. The only symmetric single-particle Hamiltonian is trivial, so following Rule 2, dynamics must be caused by interaction. In the toy model, this was achieved by applying an external magnetic field $\vec B$ that made the spin precess. To model this field, all we need is a big magnet $\cM$. Following Rule 1, this magnet must be quantum mechanical, so we represent it with a particle with maximal spin along the $x$ axis:
\begin{equation*}
\ket{M,M}_x^\cM = \frac{1}{2^M}\sum_{m=-M}^{M} \binom{2M}{M+m}^{1/2} \ket{M,m}^\cM.
\end{equation*}

We must couple this magnet to the system with a symmetric Hamiltonian, so it has to be a scalar function of $\vec J^\cM\cdot\vec\sigma^\cS$. We choose the Heisenberg coupling $H^{\cS\cM} = -2\lambda \vec J^\cM\cdot\vec\sigma^\cS$. To see that this Hamiltonian is rotationally invariant, we can write it in terms of the total angular momentum operators: $H^{\cS\cM} = -\lambda[(J^{\cS\cM})^2-(\sigma^\cS)^2 - (J^\cM)^2]$ which obviously commutes with the generators of the symmetry group. Given the system's initial state $\alpha\ket\up + \beta \ket\dn$, we can easily solve the equation of motion and get
\begin{equation}
\ket{\Psi(t)}^{\cS\cM} = \ket{M,M}_x^\cM\otimes\ket{\psi(t)}^\cS 
+C(t)\left[\frac{1}{\sqrt{2M}}\ket{M,M}_x^\cM\otimes \ket\dn^\cS + \ket{M,M-1}_x^\cM\otimes \ket\up^\cS\right]
\label{eq:state_SMt}
\end{equation}
where $\ket{\psi(t)}^\cS$ is the solution to the toy model given by Eq.~(\ref{eq:time_coeff}) with $B = \lambda(2M+1)$, and the function $C(t)$ is equal to $i\sqrt M 2 (\alpha-\beta)\sin(Bt/2)/(2M+1)$. We see that $C(t) \sim 1/\sqrt M$, so when the magnet reaches macroscopic sizes, we obtain the same formal solution as we did with the toy model. Note however that the physical description is not yet rotationally invariant as states are expressed with respect to an external $z$ quantization axis. 

Once again, what we have illustrated here with a spin-$\frac 12$ system is true in general and follows from Eq.~(\ref{eq:limitCG}) and another similar identity
\begin{equation}
\lim_{M\rightarrow\infty} \big|C(S,s+\Delta;M,M-\Delta;M+s,M+s)\big|^2 = 
\left\{
\begin{array}{l}
1 \ \ \mathrm{if}\ \Delta = 0 \\
0 \ \ \mathrm{otherwise}.
\end{array}
\right. 
\label{eq:limitCG2}
\end{equation}
Taking the limit $M \rightarrow \infty$ while keeping $2\lambda M = B$ will result in the state
\begin{equation}
\ket{\Psi(t)}^{\cS\cM} \approx \sum_s \alpha_s e^{iBst} \ket{S,s}_x^\cS \ket{M,M}_x^\cM
\label{eq:general_dynamics}
\end{equation}
as expected for a spin-$S$ particle immersed in a magnetic field along the $x$ axis. 

To eliminate the unphysical reference frame from the above discussion, we need to reintroduce the gyroscope and apply Rule 3. At this stage, only relational degrees of freedom between $\cS$, $\cM$, and $\cG$ will remain. One can easily recover the non-relational result by letting both the gyroscope and the magnet reach macroscopic sizes. Note however that we need to let the size of the gyroscope grow faster than that of the magnet. We could take for example $M^2 = G \rightarrow \infty$. After a few algebraic manipulations and keeping only the non vanishing terms, we arrive at 
\begin{equation}
\rho^{\cS\cM\cG}_{\mathrm{physical}} \approx \frac{1}{2^{2M}}\sum_{n=-M}^{M-1}\binom{2M}{M +n}
\kb{\Psi_n(t)}{\Psi_n(t)}^{\cS\cM\cG}
\label{eq:state_SMG}
\end{equation}
where we have defined
\begin{equation}
\ket{\Psi_n(t)}^{\cS\cM\cG} = \alpha(t)\ket{G+\tfrac 12+n;G+\tfrac 12}^{\cS\cM\cG} 
+ \beta(t)\sqrt{\frac{M-n}{M+n+1}}\ket{G+\tfrac 12+n;G-\tfrac 12}^{\cS\cM\cG}
\label{eq:state_n}
\end{equation}
where $\alpha$ and $\beta$ are defined in the toy model, c.f. Eq.~(\ref{eq:time_coeff}). In the above equation, the quantum numbers refer to the eigenvalues of the total angular momentum, and the joint angular momentum of the system and the gyroscope, i.e. $(J^{\cS\cM\cG})^2 \ket{a;b}^{\cS\cM\cG} = a(a+1)\ket{a;b}^{\cS\cM\cG}$ and $(J^{\cS\cG})^2 \ket{a;b}^{\cS\cM\cG} = b(b+1)\ket{a;b}^{\cS\cM\cG}$. These quantum numbers are purely relational. The quantum numbers associated to $(\sigma^\cS)^2$, $(J^\cM)^2$, and $(J^\cG)^2$ still have values $\frac 12$, $M$, and $G$ respectively, but were omitted to avoid cumbersome notation. All quantum numbers are at this stage associated to rotationally invariant observables. 

Now, observe that the binomial coefficient appearing in Eq.~(\ref{eq:state_SMG}) is sharply peaked around the value $n=0$, with a width $\Delta n \sim \sqrt M$. In this range, the term appearing under the square-root in Eq.~(\ref{eq:state_n}) is one, plus fluctuations of order $1/\sqrt M$. Following Rule 2, we conclude that with probability approaching unity as $M$ goes to infinity, the joint state of the system, magnet, and gyroscope is $\ket{\Psi_n(t)} \approx \alpha(t) \ket{G+\tfrac 12+n; G+\tfrac 12}^{\cS\cM\cG} + \beta(t) \ket{G+\tfrac 12 +n; G-\tfrac 12}^{\cS\cM\cG}$ for some random $n \in [-\sqrt{M},\sqrt{M}]$. 

The {\em reduced} state of the system and the gyroscope --- the state obtained by tracing out the relational degree of freedom involving the magnet --- is given by
\begin{eqnarray}
\rho_{\mathrm{physical}}^{\cS\cG} &\approx& 
|\alpha(t)|^2 \kb{G+\tfrac 12;G;\tfrac 12}{G+\tfrac 12;G;\tfrac 12}\nonumber \\
&+& |\beta(t)|^2 \kb{G-\tfrac 12;G;\tfrac 12}{G-\tfrac 12;G;\tfrac 12},
\label{eq:physical_SGt}
\end{eqnarray}
with interpretation that at time $t$, the system and gyroscope's spin are either parallel or antiparallel with respective probabilities $|\alpha(t)|^2$ and $|\beta(t)|^2$. There are two ways to arrive at this result. One can start from Eq.~(\ref{eq:state_SMG}), reverse the Clebsch-Gordan transformation and trace out the magnet. A more direct route is to use the fact that the map associated to tracing out a system and the map $\cE$ representing a group average as in Eq.~({\ref{eq:CPmap}) commute when the group acts unitarily on the system being traced out, i.e.
\begin{equation*}
Tr_\cB \int U^\cA(\Omega)\otimes U^\cB(\Omega) \rho^{\cA\cB} U^\cA(\Omega)^\dagger\otimes U^\cB(\Omega)^\dagger d\Omega 
= \int U^\cA(\Omega) Tr_\cB\{ \rho^{\cA\cB}\} U^\cA(\Omega)^\dagger d\Omega .
\end{equation*}
Thus, we can start directly from Eq.~(\ref{eq:state_SMt}) and trace out the magnet. Up to corrections of order $1/\sqrt{M}$, the joint state of $\cS$ and $\cG$ will be given by Eq.~(\ref{eq:state_SG}) with time dependent amplitudes, so the results of Sec.~\ref{sec:measure} apply directly, yielding Eq.~(\ref{eq:physical_SGt}).  

\subsection{Time}
\label{sec:time}

So far, we have been concerned with the rotational symmetry of physical descriptions. The remaining symmetry is time translation. The explicit time parameter $t$ appearing in the above equations is defined with respect to an unphysical reference frame, so must also be eliminated. But before getting rid of time, it is practical to build a clock! A clock $\cC$ is just a big rotating needle, so again it will be represented quantum mechanically by a spin-$C$ particle initialized in state $\ket{C,C}^\cC$. By letting this clock interact with a magnet $\cN$ (we use the letter $\cN$ for this magnet as $\cM$ is already used), it will start rotating just like a normal clock does. Of course, this clock has periodic motion --- with period $T^\cC=\pi/\lambda N$ --- so it can only keep tract of time in a fixed interval $[0,T^\cC]$.\footnote{In fact, since the gyroscope only allows us to read the clock along a single axis, the {\em observable} clock's period is really $T^\cC/2$. More sophisticated clocks could be built, but we will ignore this problem by choosing $\Lambda$ to be an even number for simplicity.}

In the real world, this problem is fixed by hooking up clocks to calendars, which break the periodicity. Here, we will circumvent this problem by assuming that the spacetime manifold (or sub-manifold of interest) has $t\in[0,T^\cC]$ with periodic boundary conditions, so our clock is well adapted. Here, periodic boundary conditions imply that as both magnets reach macroscopic size, their ratio $\Lambda = M/N$ is an integer, which is just saying that $T^\cC$ is an integer multiple of the system's precession period. 

We will eliminate time using Rule 3, exactly as we did for the rotational reference frame: we perform a group average and remove the unphysical degrees of freedom. Given a time translation operator $U(t)$, we define the TPCP map $\cT$ by the action
\begin{equation}
\cT(\rho) = \frac{1}{T^C} \int_0^{T^\cC} U(t)\rho U(t)^\dagger dt.
\end{equation}
The combined effect of $\cE$ (c.f. Eq.~(\ref{eq:CPmap})) and $\cT$ is to randomize all non physical degrees of freedom, while keeping relational ones unchanged. Unphysical degrees of freedom can then be removed from the description. 

To apply this procedure to our model, we use the the same tricks as above. Using Eq.~(\ref{eq:general_dynamics}), we obtain an expression of the time dependent state of the system, clock and two magnets. We can then trace out both magnets as both maps $\cT$ and $\cE$ act unitarily on them (in the case of $\cT$, this is only true in the asymptotic limit).  This yields the state
\begin{equation}
\rho^{\cS\cC}(t) \approx \sum_{s,s' = \pm1/2}\sum_{c,c' = -C}^C a_sa_{s'}^*\frac{1}{2^{2C}}\sqrt{\binom{2C}{C+c}\binom{2C}{C+c'}} e^{it\{B(s-s') + B'(c-c')\}} \kb{s}{s'}_x^\cS\otimes \kb{C,c}{C,c'}_x^\cC
\end{equation}
where subscripts $x$ indicate that the second quantum number refers to $J_x$, and $a_{\pm 1/2} = (\alpha \pm \beta)/\sqrt 2$. Under the periodic boundary conditions of our model, the map $\cT$ will turn the exponential into a Kronecker delta $\delta_{c+\Lambda s,c'+\Lambda s'}$, where $\Lambda$ is define above as the ratio of the two magnetic fields. We then introduce the gyroscope into the picture, and express the state in terms of the operator $\sigma_z^\cS$, $(J^{\cC\cG})^2$ and $J^{\cC\cG}_z$. Using Eq.~(\ref{eq:limitCG}), we obtain
\begin{eqnarray*}
\rho^{\cS\cC\cG} &\approx& \sum_c \sum_{s,s',r,r'} \frac{a_sa_{s'}^*}{2}
\frac{1}{2^{2C}}\sqrt{\binom{2C}{C+c+\Lambda s}\binom{2C}{C+c+\Lambda s'}} 
(-1)^{(r-1/2)(s-1/2)+(r'-1/2)(s'-1/2)}\kb{r}{r'}^\cS\\
&\otimes &\sum_{m,m'} d^C_{m,c+\Lambda s} d^C_{m',c+\Lambda s'}
\kb{G+m,G+m}{G+m',G+m'}^{\cC\cG}
\end{eqnarray*}
where again, the quantum numbers for $(J^\cC)^2$ and $(J^\cG)^2$ are constant, so were omitted. The coefficients $d^j_{m,m'} = d^j_{m,m'}(\pi/2)$ are the Wigner rotation matrices (see e.g. \cite{Sak94a}) that allow us to express $J_x$ eigenstates in terms of $J_z$ eigenstates. We can now apply the map $\cE^{\cS\cC\cG}$ to this state, and trace out the unphysical degrees of freedom to obtain
\begin{eqnarray}
\rho^{\cS\cC\cG}_{\mathrm{physical}} &\approx& \sum_c \sum_{s,s',r,r'} \frac{a_sa_{s'}^*}{2}
\frac{1}{2^{2C}}\sqrt{\binom{2C}{C+c+\Lambda s}\binom{2C}{C+c+\Lambda s'}} 
(-1)^{(r-1/2)(s-1/2)+(r'-1/2)(s'-1/2)}\nonumber\\
&\times&
\sum_{u} d^C_{u-r,c+\Lambda s} d^C_{u-r',c+\Lambda s'}
\kb{G+u;G+u-r}{G+u;G+u-r'}^{\cS\cC\cG}
\label{eq:state_SCGphysical}
\end{eqnarray}
where the two quantum numbers $a$ and $b$ appearing in the states $\ket{a;b}$ refer to the relational observables $(J^{\cC\cG})^2$ and $(J^{\cS\cC\cG})^2$ respectively.  

To ``read the time", one must measure the clock's needle orientation relative to the gyroscope; in other words, measure $(J^{\cC\cG})^2$. This will fix the value of $G+u$, and hence of $u$.  As before, this should be given the interpretation that the clock's needle is at an angle $\theta$ satisfying $\cos(\theta) = u/C$.  Just as with regular clocks, the angle of the needle $\theta$ is directly interpreted as time. We can evaluate the probability distribution $P(u)$ numerically from Eq.~(\ref{eq:state_SCGphysical}); we observe that the probability distribution for various values of $u$ is roughly given by $1/\pi\sqrt{C^2+u^2}$, see Fig.~\ref{fig:rabi} a). This will lead to a {\em flat} distribution for the values of $\theta = \cos^{-1}(u/C)$ as expected. The interpretation of this result is that when reading the clock, one gets a random answer $\theta$ taking discrete values in $[0,\pi]$ with roughly equal probabilities.

Given a value of $u$, we can obtain the conditional state of the system, clock, and gyroscope by applying the von Neumann state update rule to $\rho^{\cS\cC\cG}_{\mathrm{physical}}$ \cite{vN55a}:
\begin{equation}
\rho^{\cS\cC\cG}_{\mathrm{physical}} \xrightarrow{u} \rho^{\cS\cC\cG|u}_{\mathrm{physical}} = 
\frac{P_u \rho^{\cS\cC\cG}_{\mathrm{physical}} P_u}{P(u)}
\end{equation}
where $P_u$ is the projector onto the subspace corresponding to the measurement outcome $u$. Since here the state $\rho^{\cS\cC\cG}_{\mathrm{physical}}$ commutes with the various $P_u$, the state update rule is formally equivalent a classical Bayesian update, so it is compatible with Rule 4.

For a fixed value of clock read $\theta$, we can ask what is the orientation of the system relative to the gyroscope and clock, or in other words, measure $(J^{\cS\cC\cG})^2$. (Note that measuring the orientation of the system relative to $\cG$ or $\cC + \cG$ gives the same result when $\cG$ is of macroscopic size.) This fixes the value of $G+u+s$, and hence of $s$.  Starting from Eq.~(\ref{eq:state_SCGphysical}), we have numerically evaluated the probability of the outcome $s=-1/2$ --- the system and gyroscope's spin is antiparallel --- given a value of $\theta$ for various gyroscope sizes; results are shown on Fig.~\ref{fig:rabi}. As expected, when the size of the clock is very large (e.g. $C=400$) the results are very close to the orthodox predictions. For small clocks however, results agree for small values of $\theta$ and rapidly deteriorate. This is an interesting effect that we will discuss in Sec.~\ref{sec:decoherence}. This completes our ``translation" of the orthodox description  of a system immersed into a magnetic field into a purely relational  description.

\begin{figure}[tb] 

\center \includegraphics[width=6.5in]{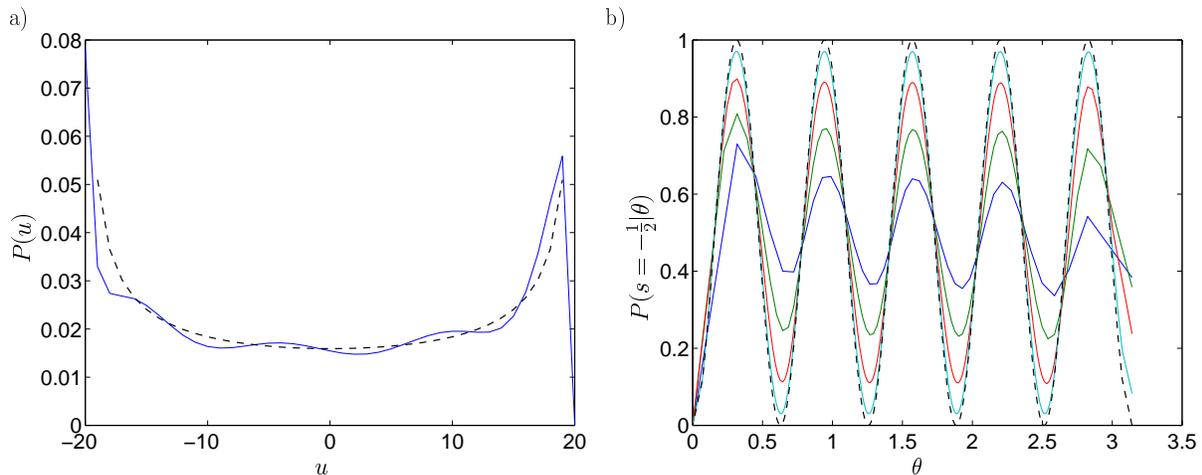}
\caption{Numerical results obtained from Eq.~(\ref{eq:state_SCGphysical}). The ``free parameters" $\alpha$ and $\beta$, reflecting the system's initial state in the non-relational theory, were fixed to $\alpha = 1$ and $\beta = 0$. The ratio $\Lambda = M/N$ of the magnet's size is fixed to 10.  a) Probability of the outcome $G+u$ for the measurement of $(J^{\cC\cG})^2$ as a function of $u$, for a clock of size $C=20$. The dash line is the function $1/\pi\sqrt{C^2-u^2}$ corresponding to a flat distribution over the value of $\theta$. b) Probability of the measurement outcome of $(J^{\cS\cC\cG})^2$ indicating an antiparallel orientation of the system and gyroscope's spins as a function of clock reading $\theta$, for various clock sizes $C = 20, 40, 100,$ and $400$. As the size of the clock increases, the result approaches the non-relational prediction $P(r=+1/2 | \theta) = |\beta(t)|^2$ given by Eq.~(\ref{eq:time_coeff}) and illustrated by the dash line on the figure. The deterioration of the result as $\theta$ increases is due to a fundamental {\em decoherence} mechanism that occurs whenever clocks are of finite size \cite{GPP04a}.}
\label{fig:rabi}
\end{figure} 

\section{Symmetries and noiseless subsystems}
\label{sec:nns}

The techniques illustrated in the previous Section were inspired by the noiseless subsystem method used to protect quantum states against undesired noise in quantum information science. The derivation we applied to the symmetry group $SO(3) \times U(1)$ representing rotation and time translation in the previous Section can be applied straightforwardly to any symmetry group acting on a finite dimensional Hilbert space. The goal of this Section is to present the general picture that naturally emerges from the four simple Rules stated in Sec.~\ref{intro}. 

Let us begin with a bit of notation. We consider a collection of quantum systems $\cS_1,\ \cS_2,\ldots$ representing for example the system of interest, a clock, a gyroscope, etc. To avoid unnecessary mathematical complications, we will assume that each of these systems is associated a finite dimensional Hilbert space $\cH^{\cS_j}$, with $dim(\cH^{\cS_j}) = d_j$. In a non-relational theory, physical states are given by rays in $\cH = \bigotimes_j \cH^{\cS_j}$. These states can be expressed in terms of arbitrary basis $\{\ket{e^1}^{\cS_j}, \ket{e^2}^{\cS_j},\ldots, \ket{e^{d_j}}^{\cS_j}\}$ which serve as a reference frame. 

\subsection{Noiseless subsystems}

A TPCP map $\cE:\cB(\cH)\rightarrow\cB(\cH)$ can be described in an operator-sum representation~\cite{Kra83a} as $\cE(\rho) = \sum_a E_a \rho E_a^\dagger$, with $\sum_a E_a^\dagger E_a = \II$ to ensure trace preservation. The algebra $\cA$ generated by the set $\{E_a,E_a^\dagger\}$ is a $\dagger$-algebra, called the {\it interaction algebra}, and as such it is unitarily equivalent to a direct sum of (possibly ``ampliated'') full matrix algebras: $ \cA \cong \bigoplus_J  \cM_{m_J}\otimes \II_{n_J}$, where $\cM_m$ is a $m$-dimensional full matrix algebra, and $\II_n$ is the $n\times n$ identity operator. This structure induces a natural decomposition of the Hilbert space
\begin{equation*}
\cH = \bigoplus_J \cH_J\otimes\cK_J,
\end{equation*}
where the ``noisy subsystems" $\cH_J$ have dimension $m_J$ and  the ``noiseless subsystems" $\cK_J$ have dimension $n_J$.

If $\cE$ is a {\em unital} quantum operation, by which we mean that
the maximally mixed state $\II$ remains unaffected by $\cE$ (i.e.,
$\cE(\II)=\II$), then the fundamental {\it noiseless subsystem}
method \cite{KLV00a,Zan01b,KBLW01a} of quantum error correction may be applied. This method makes use
of the structure of the noise commutant,
\begin{equation}
\cA' = \big\{\rho\in\cB(\cH): E \rho = \rho E \,\,\,\forall
E\in\{E_a,E_a^\dagger\} \big\},
\end{equation}
to encode states that are immune to the errors of $\cE$. Notice that with the
structure of $\cA$ given above, the noise commutant is unitarily
equivalent to $\cA'\cong \oplus_J \II_{m_J} \otimes \cM_{n_J}$.
Moreover, for unital $\cE$, the
noise commutant coincides with the fixed point set for $\cE$~\cite{BS98b,Lin99a}; i.e.,
\begin{equation*}
\cA' = Fix(\cE) = \{ \rho\in\cB(\cH): \cE(\rho)=\rho\}.
\end{equation*}
This means that a quantum state $\rho$ will not be affected by the noise operation $\cE$ if and only if it is in $\cA'$. We may thus regard the spaces $\cK_J$ as the Hilbert spaces associated to virtual particles that are not affected by the map $\cE$. The noiseless subsystem technique has recently been generalized to include non-unital maps \cite{KLP05a,NP05a}, but the unital case will be sufficient for our purposes.

\subsection{Geometrical symmetries}

We begin by considering ``geometrical" symmetries, i.e. those that are not time translation (we assume for a moment a non relativistic setting). The fundamental symmetries of the system are represented by a group $\cG$. In the example of Sec.~\ref{sec:example}, this group was $SO(3)$. The group $\cG$ acts unitarily on the state space of each system: the effect of $g \in \cG$ on the state $\ket\psi^\cS$ of system $\cS$ is represented by some unitary matrix $U^\cS(g)$. In a fundamental description, the systems should be chosen to be elementary particles, and the representations $U^{\cS_j}$ will therefore be irreducible. Thus, the different basis that can be used to express states are related to each other by a group element, i.e. any two basis $\{\ket{e^1}^{\cS}, \ket{e^2}^{\cS},\ldots, \ket{e^{d}}^{\cS}\}$ and $\{\ket{f^1}^{\cS}, \ket{f^2}^{\cS},\ldots, \ket{f^{d}}^{\cS}\}$ of $\cH^{\cS}$ are related by an element of $\cG$, in the sense that there exists a $g$ for which $\ket{f^k}^\cS = U^\cS(g) \ket{e^k}^\cS$ for all $k$, up to a permutation of the labels $k$. 

From a quantum Bayesian point of view~\cite{CFS02a,Fuc02b}, the non-relational states should be thought of as states {\em given a preferred basis} or equivalently, {\em given a preferred reference frame} $R$. Thus, we should  write $\ket\psi^{\cS_1\cS_2\ldots}_{|R}$ for states expressed with respect to the reference frame $R$. The same physical state can be expressed in terms of an other reference frame $R'$ as $\ket\psi^{\cS_1\cS_2\ldots}_{|R'} = U^{\cS_1}(g)\otimes U^{\cS_2}(g)\otimes\ldots \ket\psi^{\cS_1\cS_2\ldots}_{|R}$, where $g$ if the group element relating the basis associated to $R$ and $R'$. But background independence tells us that $R$ doesn't exist. Following the Bayesian prescription, in the absence of an external reference frame, the state assigned to the collection of system should be a statistical mixture of  the $\ket\psi^{\cS_1\cS_2\ldots}_{|R}$ averaged over all reference frames. This leads to 
\begin{eqnarray}
\rho_{\rm physical}^{\cS_1\cS_2\ldots} &=& \cE^{\cS_1\cS_2\ldots}(\kb\psi\psi^{\cS_1\cS_2\ldots}_{|R}) \\
&=& \int_\cG U^{\cS_1}(g)\otimes U^{\cS_2}(g) \otimes \ldots 
\kb\psi\psi^{\cS_1\cS_2\ldots}_{|R} U^{\cS_1}(g)^\dagger\otimes U^{\cS_2}(g)^\dagger \otimes \ldots dg
\end{eqnarray} 
where $dg$ is the group invariant measure satisfying $\int_\cG dg =1$. This choice of ``flat" distribution reflects our complete ignorance of a preferred reference frame $R$, so it is well justified in a Bayesian approach. This defines a TPCP map $\cE^{\cS_1\cS_2\ldots}$ analogue to the one defined at Eq.~(\ref{eq:CPmap}) in our example of Sec.~\ref{sec:example}. This averaging procedure is the exact analog of the rule $P(a) = \sum_b P(a|b)P(b)$ of classical probability theory, relating the probability of event $a$ to the conditional probability of $a$ given a value of $b$ and the prior probability of $b$. 

This Bayesian inspired group average differs from the ``coherent" group average $\int _\cG U^{\cS_1}(g)\otimes U^{\cS_2}(g)\otimes\ldots \ket{\psi}^{\cS_1\cS_2\ldots} dg$ commonly encountered in quantum gravity (see \cite{Mar00a} and references therein). In the case of rotational symmetry for example, the coherent group average simply projects onto the spin-zero subspace. This could annihilate  the state, e.g. if the ``universe" contained an odd  number of particles with half odd integer spin. A clear advantage of the ``statistical" group average used here is that it is trace preserving. Moreover, carrying the group average at the level of $\cB(\cH)$ rather than $\cH$ may eliminate some mathematical complications that arise when the symmetry group only admits projective representations, i.e. when the left and right invariant Haar measure differ.

The map $\cE^{\cS_1\cS_2\ldots}$ constructed from tensor products of irreducible representations of the symmetry group induces a partition of the total Hilbert space 
\begin{equation}
\cH = \bigotimes_j \cH^{\cS_j} = \bigoplus_J \cH_J \otimes \cK_J
\end{equation}
where $J$ is a label for the different unitary irreducible representations of $\cG$, $\cH_J$ is the sector on which the $J$th representation acts, and $\cK_J$ is the space associated to the degeneracy of the $J$th representation. By virtue of Schur's lemma, the effect of the map $\cE^{\cS_1\cS_2\ldots}$ can easily be described in terms of this decomposition:
\begin{equation}
\cE^{\cS_1\cS_2\ldots}(\rho) = \sum_J \frac{\II_{\cH_J}}{m_J}\otimes Tr_{\cH_J}\{P_J \rho P_J\}
\label{eq:effect}
\end{equation}
for all $\rho \in \cB(\cH)$. The projectors $P_J$ are defined by $P_J \cH = \cH_J\otimes \cK_J$, $\II_{\cH_J}$ is the identity operator on $\cH_J$, and $m_J = dim(\cH_J)$. The operation $Tr_{\cH_J}: \cB(\cH_J\otimes\cK_J) \rightarrow \cB(\cK_J)$ denotes the partial trace. In words, this map first imposes a superselection rule forbidding coherent superpositions across different $J$ sectors. Then, within each superselected sector, it completely randomizes the state over the $\cH_J$ sector. 

At this stage, the analogy with noiseless subsystems is clear. Losing an external reference frame induces some kind of  ``noise" into our physical description. We can think of each sectors $\cK_J$ as virtual {\em subsystems} that are immune to this noise. Obviously, these sectors must encode only {\em relational} information --- information that is independent of any external reference frame.  On the other hand, the sectors $\cH_J$ contain no information whatsoever about the physical system as they are always in a maximally mixed state in the absence of a reference frame. Thus, we can drop these ``noisy" sector an simply write $\rho_{\rm physical} = \sum_J p_J \rho_J$ where the $p_J$ were introduced so that the $\rho_J \in \cB(\cK_J)$ have unit trace.\footnote{One has to be a bit careful with this notation as the $\rho_J$ do not necessarily have the same dimension, as they act on different superselected sectors.} The interpretation of this state follows straightforwardly from Rule 4: the system is in one of the states $\rho_J$ with respective probability $p_J$.  

We note that, since each symmetry of the system is associated a conserved quantity, the superselection induced by the loss of an external reference frame implies superselection of conserved quantities. Hence, these quantities will always behave ``classically", as was noted before e.g.  in the history formulation of quantum theory \cite{HLM95a}.  

\subsection{Time translational symmetry}

Time translational symmetry is treated in a completely analogous fashion. According to Rule 2, the Hamiltonian $H$ has all the symmetries of the system; in other words, it should be time independent and invariant under the action of $\cG$. A time-dependent quantum states should be thought of as states {\em given an external clock} $C$, and should accordingly be denoted $\ket{\psi(t)}_{|C}$. If $C$ and $C'$ are two clocks with associated time coordinate $t$ and $t'$, we have $\ket{\psi(t')}_{|C'} = e^{iH(t-t')}\ket{\psi(t)}_{|C}$. Once again following the Bayesian prescription, the absence of such an external clock leads to time averaging 
\begin{equation}
\rho_{\rm physical} = \cT(\kb\psi\psi_{|C}) = \frac 1T \int_0^T e^{-iHt} \kb\psi\psi_{|C} e^{iHt} dt
\end{equation}
where $dt$ is the time translational invariant measure satisfying $\int_0^T dt = T$, and $T$ is the period of the Hamiltonian $H$. The effect of $\cT$ will be to impose an energy superselection rule, so 
\begin{equation}
[\rho_{\rm physical},H] = 0.
\label{eq:WdW}
\end{equation} 
This may appear awkward since, in a non-relational framework, this commutator generates the system's dynamics. But in a relational theory, the Hamiltonian naturally leads to a {\em constraint} of some sort. For example, the Wheeler-DeWitt equation \cite{dW67a} $H\ket\psi =0$ is just a special case of Eq.~(\ref{eq:WdW}). 

Since $H$ has the system's symmetries, it commutes with the elements of $\cG$. Hence, the effect of the map $\cT$ will be to break up each of the sectors $\cK_J$ imposed by the geometrical symmetries into further noisy and noiseless sectors. One can use a more direct route and treat $\cG \times U(1) = \{g \circ t:\  g\in \cG,\ t\in [0,T]\}$ as {\em the} symmetry group $\cG'$ of the system and apply the noiseless subsystem techniques directly to the interaction algebra generated by $\cG'$. 

The end product of this procedure is a relational quantum state with one quantum number that is given the interpretation of time and that is {\em classically} correlated with the other quantum numbers. This so called {\em relational time} was first suggested by Page and Wootters~\cite{PW83b}. In fact, these authors have described two distinct mechanisms by which dynamics could arise from a stationary state: either through quantum or classical correlations between the clock and the other degrees of freedom. When the state of the entire universe is {\em pure}, quantum correlations (or entanglement) are the only correlations available, so dynamics will unavoidably be caused by entanglement. This will be the case for example when considering the Wheeler-DeWitt equation. This type of clock has been investigated by many~\cite{UW89a, Peg91a,GP01b}. However, Equation~(\ref{eq:WdW}) admits mixed state solutions, and we arrived at mixed state description following an arguably reasonable set of Rules (particularly from a Bayesian perspective) and plain quantum mechanics. This might be a hint that time arises through {\em classical} correlations rather than entanglement, and that the Hamiltonian constraint equation should be imposed at the level of $\cB(\cH)$ rather than $\cH$. These distinctions between our construction and what is customarily assumed can be traced back to the use of a statistical rather than a coherent group average.

\section{Discussion}
\label{sec:discussion}

\subsection{Fundamental decoherence}
\label{sec:decoherence}

We see on Fig.~\ref{fig:rabi} b) that the relational theory does not reproduce the orthodox predictions exactly. In fact, the prediction become worst as the clock time parameter $\theta$ increases: curves are closer to the dash curve at $\theta = 0$ than they are at $\theta = \pi$. Moreover, these effect decrease as the size of the clock increases, but are always present for finite clocks. This is due to the fact that time is now a quantum variable, and as such it is subject to quantum fluctuations. The effect of such a diffuse time has been investigated as a fundamental decoherence mechanism~\cite{GPP04a,GPP05a,Mil03a,MP05a}, taking pure states into mixed states~\cite{Zur03a}. 

Again, the Bayesian approach helps understanding the origin of this fundamental decoherence. In the absence of quantum fluctuation of the clock variable, the state of the systems conditioned on the clock reading $t$ is given by $\rho(t) = e^{-iHt}\rho(0)e^{iHt}$. However, the reading of a finite dimensional clock yields a diffuse time value: time is known within a finite accuracy. Thus, given a clock reading ``$\theta$'', the state of the system should be
\begin{equation}
\rho(\theta) = \int P(t|\theta) e^{-iHt} \rho(0) e^{iHt} dt
\end{equation}
where $P(t|\theta)$ represents our {\em a posteriori} probability distribution over the value of $t$ given our knowledge of $\theta$. The decoherence rate will be directly related to the width of this distribution \cite{Mil03a,MP05a}, which in turn is a function of the clock's size. Determining optimal tradeoff between clock-size and decoherence rate in a relational theory is obviously an interesting question and we leave it for future investigation. Moreover, the finite-size effect of the other systems $\cG$, $\cM$, and $\cN$ --- which we have assumed to be infinite in our numerical analysis --- will add on to this disagreement between orthodox and relational predictions. For example, Eq.~(\ref{eq:rho_phys}) leads to slightly modified probability rule when the gyroscope is of finite size. 

It is interesting to note that this effect introduces an arrow of time: even though the relational theory is time-symmetric (or more precisely time-independent), it yields predictions that are not symmetric with respect to the clock time, as is clearly illustrated on Fig.~\ref{fig:rabi} b). This suggests that the arrow of time could emerge from the finite size of our clocks, and that this could be verified experimentally. Recall however that our setting only allows us to measure the clock's spin along a single axis, and as a consequence we cannot distinguish the time range $\theta \in [0,\pi]$ from $\theta \in [\pi,2\pi]$. Being able to distinguish the time range $\theta \in [\pi,2\pi]$, e.g. by introducing a second gyroscope aligned along the $y$ axis, we would observe a ``recoherence" phase during the second half of the universe's period. Indeed, this follows from the periodic boundary conditions: the amplitude of the oscillation at time $\theta = 2\pi$ is equal to the amplitude at time $\theta = 0$. Such recoherence can only occur if $\partial \rho^{\cS\cG} / \partial \theta$ --- the derivative of the system's state with respect to the clock time $\theta$ --- is non-local in time; if the equation of motion has a memory term. In a Markovian approximation where we neglect the memory term, the recoherence phase will disappear, leading to an effective arrow of time. A detailed study of this effect is left to future investigation.

\subsection{Spin networks}
\label{sec:SN}

We will now revisit our toy model and introduce a diagrammatic representation for every step that went into  our calculations. The five quantum systems $\cS,\ \cM,\ \cC,\ \cN$ and $\cG$ are each assigned a angular momentum operator $\vec J^\cS,\ \vec J^\cM,\ \vec J^\cC,\ \vec J^\cN$, and $\vec J^\cG$ respectively. The first step to solve the non-relational dynamical equations was to express the state of $\cS$ and $\cM$ in terms of their total angular momentum operator, and similarly for the the pair $\cC-\cN$.  Thus, we define two new operators $\vec J_1$ and $\vec J_2$ satisfying $\vec J^\cS + \vec J^\cM + \vec J_1 = 0$ and $\vec J^\cC  + \vec J^\cN - \vec J_2 = 0$ (the signs might appear arbitrary but are necessary). We can represent this graphically as follows:
\begin{equation}
\includegraphics{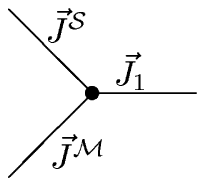}
\end{equation}
and similarly for the $-\vec J^\cC$, $-\vec J^\cN$, $\vec J_2$ triplet. 

In the next step, we combined the angular momentum of the gyroscope to $\vec J_2$ in order to ``read the time". This defines a new operator $\vec J_3$ satisfying the relation $\vec J^\cG + \vec J_2 + \vec J_3 = 0$, or in other words $\vec J_3 = -(\vec J^\cC + \vec J^\cN + \vec J^\cG)$. Finally, to measure the system's state relative to the gyroscope and clock, we have defined yet an other operator $\vec J_{total}$ satisfying the relation $\vec J_1 + \vec J_3 + \vec J_{total} = 0$, or equivalently $\vec J_{total} =  \vec J^\cS + \vec J^\cM + \vec J^\cC + \vec J^\cN + \vec J^\cG$. Combining all these steps together yields the graph
\begin{equation}
\includegraphics{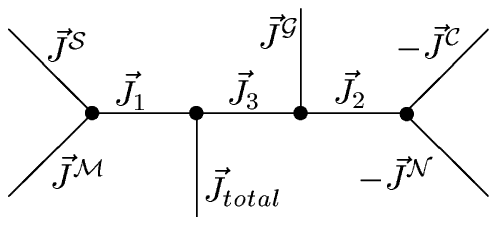}
\label{eq:graph2}
\end{equation}

At this stage, we get rid of the directional reference frame by performing a group average over the symmetry group $SO(3)$ and eliminate the non-relational (or noisy) degrees of freedom. In the diagrammatic representation, this essentially boils down to removing the arrows from the operators! Hence, we are going to replace each operator by its $j$ value, i.e. perform the substitution $\vec J \rightarrow j$ such that $\vec J^2 = j(j+1)$. However, not all edges of the graph have a fixed value of $j$, so we will need to introduce superpositions of the graph with different values of $j$. The $j$ values associated to the five systems were fixed from the onset. The $j$ value associated to the $\vec J_{total}$ needs not to take a definite value, but it is superselected due to the groupe averaging procedure, and as so, it can only be in classical statistical mixtures of different $j$ values. The graph we obtain is therefore
\begin{equation}
\includegraphics{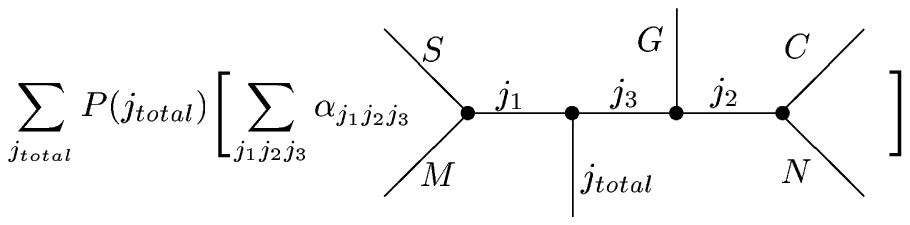}
\label{eq:graph3}
\end{equation}
where we have introduced the ``[ ]" notation as a shorthand for a convex combination of the projectors associated to the states inside the brackets, i.e. $\sum_a p_a [\ket{\Gamma_a}] = \sum_a p_a \kb{\Gamma_a}{\Gamma_a}$. The various coefficients $\alpha_{j_1j_2j_3}$ and $p_{j_{tot}}$ can all be worked out from the initial conditions of all five systems in the non-relational theory $\ket{\Psi(0)}^{\cS\cM\cC\cN\cG} = \sum_{smcng} \beta_{smcng} \ket{S,s;M,n;C,c;N,n;G,g}$. The coefficients $\alpha_{j_1j_2j_3}$ and $P(j_{total})$ appearing in Eq.~(\ref{eq:graph3}) will be respectively linear and quadratic combinations of the $\beta_{smcng}$ with appropriate Clebsch-Gordan coefficients.

The last step of our construction is to apply the time averaging map $\cT$, which leads to energy superselection. The total Hamiltonian can be expressed in terms of the $j$ values of the graph $H = \lambda[ j_1(j_1+1)+j_2(j_2+1) - S(S+1)-M(M+1)-C(C+1)-N(N+1)] $. Since $S,\ M,\ C,$ and $N$ are fixed constants, the independence of the theory on an external clock implies superselection of  $j_1(j_1+1)+j_2(j_2+1)$, which can be imposed by adding a Kronecker delta in Eq.~(\ref{eq:graph3}). Hence, we obtain the purely relational background independent state
\begin{equation}
\includegraphics{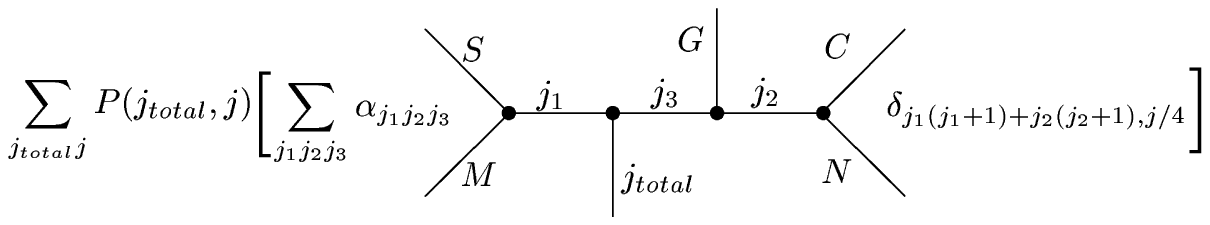}
\end{equation}

The decorated graphs $\Gamma$ --- called {\em spin networks} \cite{Pen71a,RS95a,Baez95a,Majo00a}--- corresponds to basis states $\ket\Gamma$ for the relational theory's Hilbert space $\cH_{rel}$. In our model, the free edges must have a fixed $j$ value, or a statistical mixture of such values, while the interior edges can be in quantum superpositions. A vertex with incoming edges labeled $j_1$, $j_2$, and $j_3$ is associated an {\em intertwining operator}, that is a map $\mathbb C^{2j_1+1}\otimes \mathbb C^{2j_2+1}\otimes \mathbb C^{2j_3+1} \rightarrow \mathbb C$. This map simply gives the Clebsch-Gordan coefficients needed to go from Eq.(\ref{eq:graph2}) to Eq.(\ref{eq:graph3}).  The map $\cT$ is a ``sum over histories" of the graphs, e.g. $\cT(\rho) = \int U(t)\rho U(t)^\dagger dt$. In the loop approach to quantum gravity, this sum is performed using {\em spin foams} (see e.g. \cite{Baez00a}) which are analog to Feynman diagrams used in quantum electrodynamics. However, the sum over histories is here carried at the level of $\cB(\cH_{rel})$, not on $\cH_{rel}$ as it is usually the case; it is not a coherent sum. Again, this is a consequence of the Hamiltonian constraint $[\rho,H]=0$ that naturally arises in our theory and differs from the usual equation $H\ket\psi = 0$. 

The fact that spin networks can serve as basis states in our relational model is not very surprising since they can be used to describe gauge independent observables in Yang-Mills theories, e.g. generalized Wilson loop operators \cite{KS75a,Baez96a}. Nevertheless, we believe that there is a lot to be learned about the low energy limit of spin foams models of quantum gravity from this simple analogy. In the absence of experimental guidance, connexions with well established physical regimes of the theory can be a quite useful. Spin networks are likely to play an important role in quantum gravity, but their low energy limit is poorly understood. Deriving them from textbook quantum mechanics combined with arguably reasonable Rules can thus yield interesting insights. One could of course attempt to repeat the construction with a different (and undoubtedly more interesting) symmetry group with the help of the studies pursued in the context of spin networks (e.g. see \cite{FL03a}), but this is beyond the scope of this paper. 

On the other hand, our construction also shows how the tools developed in quantum gravity can be useful in quantum information science. In particular, we can consider the group $\cG$ generated by the interaction algebra $\cA$ of a collective TPCP map $\cE$. Then, we can construct spin networks decorated by irreducible representation of $\cG$, just like we did here for the group $SU(2)$. The state space of the {\em noiseless subsystems} will thus correspond to the intertwiner space of the spin network~\cite{GL05a}. 

\subsection{Other symmetries}
\label{sec:other_symmetries}

We conclude this discussion on a more speculative tone. The basic idea exploited over and over in the present paper is to eliminate non-relational degrees of freedom by performing averages over the symmetry group of the system. This technique is well justified from a Bayesian point of view and relies on the tools of quantum information science. As mentioned at the end of the last Section, the next logical step will be to apply the technique to more interesting symmetry groups. This program should be tractable for the Lorentz group of a free field \cite{Tol97a,Maz00a} since quantum field theory is already ideally set up for this purpose, with the different particles corresponding to different irreducible representations of the symmetry group. The non-compactness of the Lorentz group certainly create extra mathematical complications, but can nonetheless be handled in principle~\cite{FL03a}. A system of harmonic oscillators also has a non-compact symmetry group, yet our program applies almost straightforwardly to this case~\cite{MP05a}. For an interacting field however, the group average becomes very difficult to perform; one can instead try to formulate the theory directly in terms of invariant observables (as in algebraic quantum field theory~\cite{Haa92a}).  

One should also attempt to extend the program to local symmetries, i.e. gauge groups. A lattice Abelian gauge theory would be a good place to start. In this context, gauge fixing is equivalent to introducing a non-physical ``reference frame" for the gauge field. Hence, our construction would eliminate the gauge degree of freedom, and the emerging relational quantum theory should not contain any gauge field; it should be expressed entirely in terms of Wilson loop like observables --- hence with spin networks~\cite{KS75a,Baez96a}. 

Finally, the ultimate goal is of course the diffeomorphism group. Our techniques cannot be applied in this setting for the simple reason that there is no non-relational theory that it can be applied to. Instead, the idea behind the loop approach to quantum gravity is to construct a theory that is diffeomorphism invariant --- and hence relational --- from the onset. 

\section{Conclusion}
\label{sec:conslusion}

In this paper, we have argued that when described at a fundamental level, i.e. in the absence of semi-classical approximations, quantum theory is naturally relational. Quantum states are independent of any external reference frame or clock, but only describe relations between physical systems. A non-relational description can be recovered as a semi-classical approximation of the relational theory. We have illustrated this thesis with the help of a collection of particles with spin --- which we use to model a clock, a gyroscope, magnets, etc. --- and discussed how the idea carries over to general quantum systems. Our construction was motivated by a Bayesian approach to quantum mechanics and borrowed tools from quantum information science.  Using elementary quantum mechanics, we have re-derive some well known concepts. By treating time as a quantum mechanical system --- a clock --- we were naturally lead to the notion of relational time of Page and Wootters~\cite{PW83b}. Quantum fluctuations of the clock variable predicts a fundamental decoherence mechanism, recently discussed by Gambini, Porto and Pullin~\cite{GPP04a}, and clearly seen in our numerical analysis. As a consequence, an effective arrow of time emerges from a time-independent theory. Finally, basis states for the relational theory can be described in terms of spin-network introduced by Penrose~\cite{Pen71a} and extensively studied in loop quantum gravity~\cite{Baez95a,RS95a}. 

However, these concepts emerged from our construction in a slightly different way than they usually do, and this might lead to interesting physical insights. The most important distinction concerns the Hamiltonian constraint: we found that physical states must commute with the Hamiltonian, while it is usually assumed that physical states must be annihilated by the Hamiltonian. This distinction is very important, specially when considering mixed states solution that are naturally expected in a Bayesian approach. As a consequence, we found that relational time rises from {\em classical} correlations between the clock and system of interest, not entanglement as it is usually assumed. This distinction also suggests that the sum over histories associated to the Hamiltonian constraint --- implemented with spin foams in loop quantum gravity --- should be carried at the level of {\em operators} rather that vectors of the Hilbert space. It is an open question whether this statistical group average can be consistently and formally incorporated into a generally covariant theory, and if so, whether it leads to physical predictions that differ from those obtained through coherent group average. Since quantum gravity is difficult to probe experimentally, deriving some of its concepts from familiar and overwhelmingly tested regimes of quantum theory as we did in this paper may offer valuable physical guidance. 

Finally, our study illustrates the usefulness of quantum information science --- and particularly the Bayesian view ---  in quantum gravity, Wheeler's ``it from bit"~\cite{Whe91a}; a point of view that has been gaining popularity lately, e.g. \cite{GL05a,Llo05a}. Conversely, we have also exposed how the tools developed in a background independent theory such as loop quantum gravity have direct applications in quantum information: e.g. by demonstrating that the intertwiners form a basis for the noiseless subsystem of a collective noise operation, all the mathematical baggage of spin networks carries over to the study of noiseless subsystems. 

\medskip
\noindent {\em Acknowledgements} --- We thank Gerard Milburn and Kenny Pregnell for stimulating and enjoyable discussions on relational quantum theory and helpful comments on this manuscript, and Michael Nielsen for a useful tip on presentation. We also acknowledge Fotini Markopoulou for interesting exchanges on spin networks and quantum information. Finally, we thank Florian Girelli, Don Marolf, Lee Smolin, and Rafael Sorkin for helpful discussions that followed a talk based on the present manuscript.


\end{document}